\documentclass{elsart1p}
 \usepackage{graphicx}
 \usepackage{epsfig}

\usepackage{amssymb}

\begin{document}

\begin{frontmatter}

\title{Relativistic Three-Body Scattering in a First Order Faddeev Formulation}

\author{Ch. Elster$^{(a)}$, T. Lin$^{(a)}$, W. N. Polyzou$^{(b)}$, W.
Gl\"ockle$^{(c)}$}

\address{(a) Department of Physics and Astronomy,  Ohio University, Athens, OH
45701, USA \\ (b) Department of Physics and Astronomy, The University of Iowa, Iowa
City, IA 52242, USA \\ (c) Institute for Theoretical Physics II, Ruhr-University Bochum,
D-44780 Bochum, Germany}

\begin{abstract}
Relativistic Faddeev equations for three-body scattering at arbitrary
energies are solved in first order in the two-body transition operator in
terms of momentum vectors without employing a partial wave decomposition.
Relativistic invariance is incorporated within the framework of Poincar{\'e}
invariant quantum mechanics. Based on a Malfliet-Tjon type interaction,
observables for elastic and breakup scattering are calculated and compared
to the nonrelativistic ones.
\end{abstract}
\end{frontmatter}

Traditionally three-nucleon calculations are carried out by solving Faddeev
equations in a partial wave truncated basis, working either in momentum or
in coordinate space. In Ref.~\cite{Liu:2004tv} the Faddeev equations were
solved directly as function of vector variables for scattering at intermediate 
energies. A key advantage of this formulation lies in its applicability at
higher energies, where the number of partial waves proliferates. 
 We investigate relativistic three-boson scattering in the framework of 
Poincar\'{e} invariant quantum mechanics. 
The main points are the construction of unitary irreducible
representations of the Poincar\'{e} group, both for noninteracting and
interacting particles.
The dynamics is generated by a Hamiltonian, and the equations we use have
the same operator form as their nonrelativistic counterparts, however the
ingredients are quite different. 

In the following we concentrate on the leading-order term of the Faddeev
multiple scattering series within the framework of Poincar{\'e} invariant
quantum mechanics. A detailed description of the formulation can be found in
Ref.~\cite{Lin:2007ck}. As a simplification we we consider three-body
scattering with spin-independent interactions of Yukawa type, which is
mathematically equivalent to three-boson scattering. The interaction employed
is of Malfliet-Tjon type, i.e. consists of a short range repulsive and
intermediate range attractive Yukawa force with parameters chosen such that a
bound state at $E_d=-2.23$~MeV is supported~\cite{Lin:2007ck}. 
To obtain a valid estimate of the size of relativistic effects, it is
important that the interactions employed in the relativistic and
nonrelativistic calculations are phase-shift equivalent. We follow here the
suggestion by Coester, Piper, and Serduke (CPS) in constructing a 
phase equivalent interaction from a non-relativistic 2N
interaction~\cite{CPS}.
In the CPS method the relativistic interaction can not be analytically
calculated from the non-relativistic one. However, there is a simple analytic
connection between the relativistic and non-relativistic two-body t-matrices
\begin{equation}
\nonumber
t_{re}({\bf p},{\bf p}';2E^{rel}_p) =
\frac{2m}{\sqrt{m^2+p^2}+\sqrt{m^2+p'^2}} \; t_{nr}({\bf p},{\bf
p}';2E^{nr}_p),  
\label{cps}
\end{equation}
where $2E^{rel}_p = 2\sqrt{m^2+p^2}$ and $2E^{nr}_p = \frac{p^2}{m}+2m$.
This relativistic two-body t-matrix $t_{re}({\bf p},{\bf p}';2E^{rel}_p)$ is
scattering equivalent to the non-relativistic one at the same relative
momentum {\bf p}~\cite{Keister:2005eq}. This t-matrix serves then as input to
obtain the Poincar{\'e} invariant transition amplitude of the 2N subsystem
embedded in the three-particle Hilbert space via a first resolvent method
 as layed out in Ref.~\cite{Lin:2007ck}.

\begin{figure}[t]
\begin{center}
\includegraphics*[width=5.9cm,height=4.2cm]{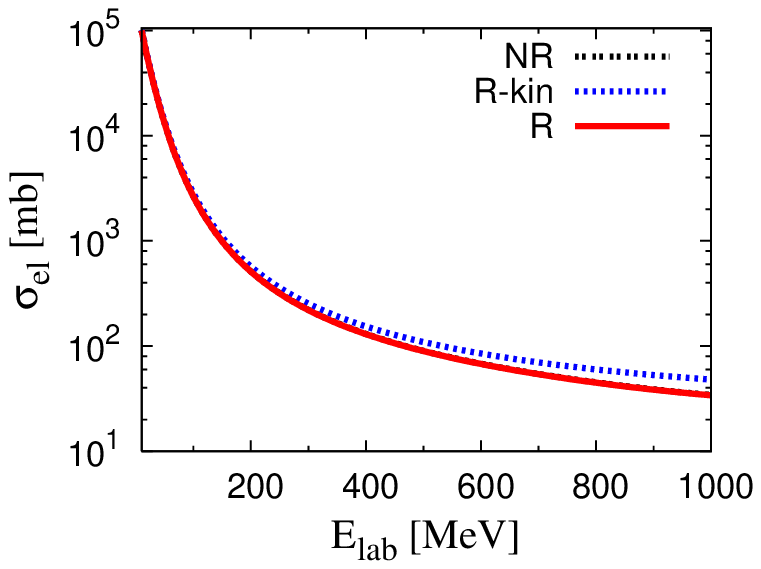} 
\includegraphics*[width=5.9cm,height=4.2cm]{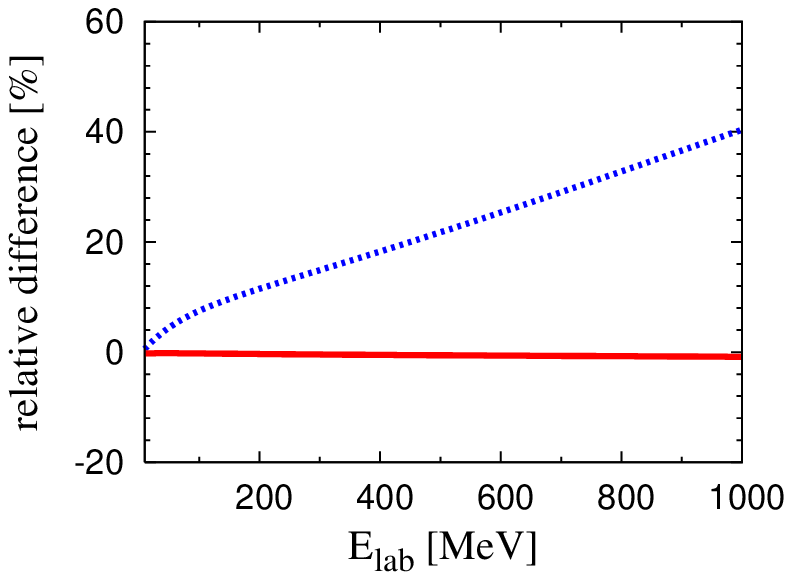}
\end{center}
\caption{The total c.m. cross section for elastic scattering calculated from
a Malfliet-Tjon type potential (left panel). The nonrelativistic calculation
is given by the dash-dotted line (NR), the relativistic one by the solid
line (R). The dashed line (R-kin) represents a calculation which only contains the
effects of relativistic kinematics. The right panel shows the relative
difference of the two relativistic calculations with respect to the
nonrelativistic one.
} 
\end{figure} 

\begin{figure}[b]
\includegraphics*[width=6.5cm,height=4.2cm]{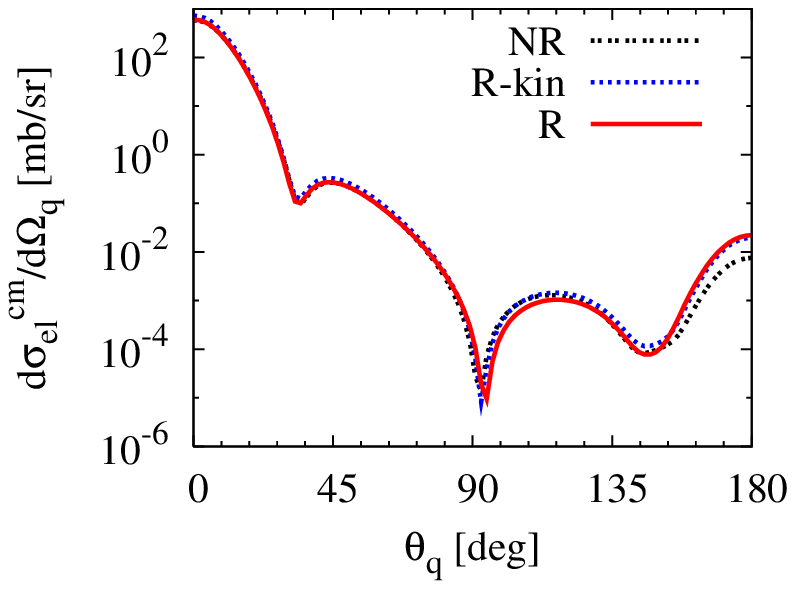}
\includegraphics*[width=6.5cm,height=4.2cm]{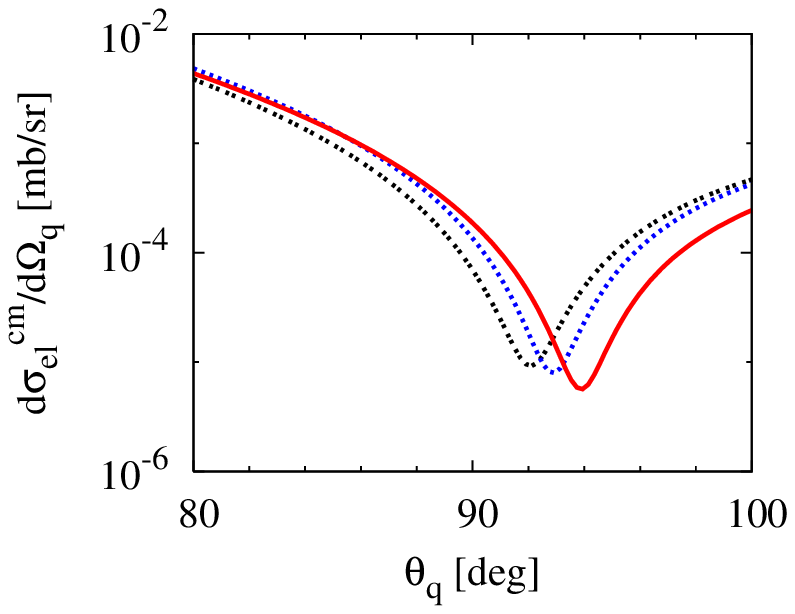}
\caption{The differential cross section for elastic scattering at 0.5 GeV
projectile energy as function of the laboratory scattering angle. 
The definition of the lines is the same as in Fig.~1. 
}
\end{figure}

In Fig.~1 we consider the total cross section for elastic scattering,
$\sigma_{el}$ for projectile kinetic energies from 10~MeV up to 1~GeV.
Starting from the non-relativistic cross section, we successively implement
relativistic features to study them in detail.  The dashed line labeled {\it
R-kin} shows the cross section when only the effects due to relativistic
kinematics, like relativistic transformations from laboratory to c.m. frame,
phase-space factor, and the relativistic kinematics due to the
Poincar{\'e}-Jacobi coordinates are taken into consideration. Implementing in
addition the effects due to the relativistic dynamics, i.e. solve the first
order equation with the 2N amplitude embedded in the three-particle Hilbert
space exactly, leads to a result close to the non-relativistic one. This can be
taken as evidence that in a first order calculation the relativistic effects
are mostly contained in the 2N amplitude, and if both, the relativistic and
non-relativistic ones are equivalent, both 3N cross sections should be very
close. Additional effects should be expected  
from the full solution of the
Faddeev equation. The differential cross section for 0.5~GeV laboratory
projectile energy is shown in Fig.~2. At the backward angle, the highest
momentum transfer, the difference between the relativistic and
non-relativistic calculations is clearly largest. A close-up of the second
minimum shows that it is shifted by relativistic kinematic effects towards
larger angles, and the relativistic dynamics pushes it even further out.  	
As an example of breakup processes we consider inclusive breakup scattering at
0.495~GeV in Fig.~3. Very obvious is the shift of the quasi-free scattering
(QFS) peak towards smaller ejectile energies, once relativistic kinematics is
introduced into the calculation. The relativistic dynamics only influences
the height of the peak, not its position. Since this is first order
calculation, the peak height may be influenced in addition by multiple
scattering contributions. A comparison of the two angles shown in Fig.~3 also
illustrates, that relativistic effects can be quite different in different
configurations: while the peak position always shifts towards smaller
ejectile energies, the peak height can either be decreased (18$^o$) or
increased (24$^o$). 

\begin{figure}[t]
\begin{center}
\includegraphics*[width=6.0cm,height=4.0cm]{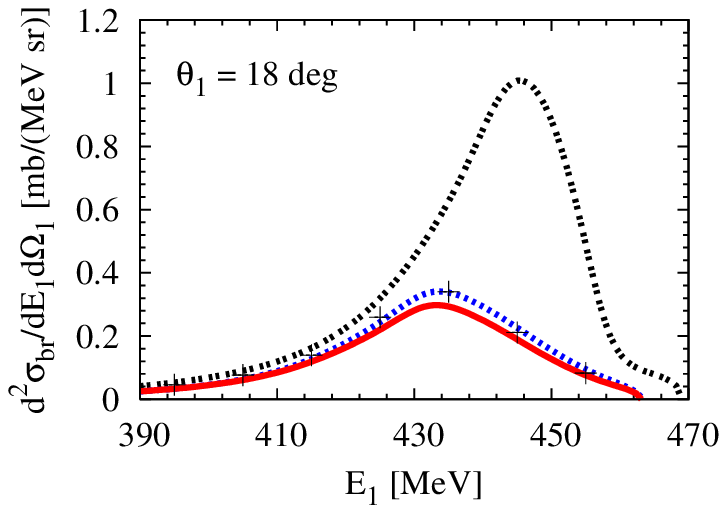}
\includegraphics*[width=6.0cm,height=4.0cm]{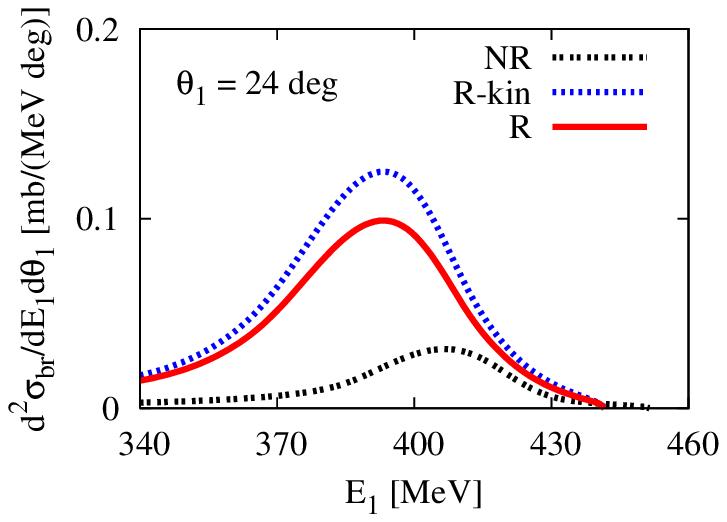}
\end{center}
\caption{The inclusive cross section at 0.495~GeV laboratory projectile
kinetic energy as function of the energy of the emitted particle and fixed
emission angles of 18$^o$ and 24$^o$ degrees. The definition of the lines is
the same as in Fig.~1. The data are from Ref.~\protect\cite{chen}
}
\end{figure}

\end{document}